\input{aipcheck}

\documentclass[
    ,final            
  ]
  {aipproc}

\layoutstyle{6x9}

\begin{document}

\title{Algorithms for Faster and Larger Dynamic Metropolis Simulations}

\author{M.A.\ Novotny}{
  address={
Dept.\ of Physics and Astronomy, ERC Center for Computational Sciences,
P.O.\ Box 5167,
Mississippi State University,
Mississippi State,
MS 39759-5167
}
}

\iftrue
\author{Alice K.\ Kolakowska}{
  address={
Dept.\ of Physics and Astronomy, ERC Center for Computational Sciences,
P.O.\ Box 5167,
Mississippi State University,
Mississippi State,
MS 39759-5167
}
}

\author{G.\ Korniss}{
  address={
Dept.\ of Physics, Applied Physics, and Astronomy,
Rensselaer Polytechnic Institute,
110 8$^{th}$ Street,
Troy, NY 12180-3580
}
}

\begin{abstract}
In dynamic Monte Carlo simulations, using for example the 
Metropolis dynamic, it is often required to simulate for 
long times and to simulate large systems.  We present an 
overview of advanced algorithms to simulate for longer times 
and to simulate larger systems.  The longer-time algorithm 
focused on is 
the Monte Carlo with Absorbing Markov Chains (MCAMC) algorithm.  It 
is applied to metastability of an Ising model on a small-world network.
Simulations of larger systems often 
require the use of non-trivial parallelization.  
Non-trivial parallelization of 
dynamic Monte Carlo is shown to allow scalable algorithms, and
the theoretical efficiency of such algorithms are described.  
\end{abstract}

\maketitle


\section{Introduction}

Dynamic Monte Carlo is used when dynamic information about a 
particular system is required.  
For example, for 
spin-1/2 lattice systems, starting from a quantum Hamiltonian 
coupled to a heat bath, the underlying dynamic for the Ising 
model can be derived as: 1) randomly and uniformly choose one 
spin; 2) decide whether or not to flip the spin based on a 
spin-flip probability $p$.  The functional form for $p$ may 
for instance be
Metropolis \cite{METR53}, Glauber (derivable from coupling the quantum 
system to a fermionic heat bath \cite{MART77}), 
or a form obtained from coupling the quantum system 
to a bosonic heat bath \cite{PARK02}. Since the simulated dynamic 
is defined by the underlying physical system, it should not be altered.
While remaining faithful to the dynamic, algorithms that allow for 
long-time simulations and non-trivial parallelization are 
still possible.  
Some of these algorithms will be presented (for a review see 
\cite{NOVO01}).  

In this article we review the use of the Monte Carlo with 
Absorbing Markov Chains (MCAMC) method and apply the method to an 
Ising ferromagnet on a small-world network.
We also describe the use of ideas from 
non-equilibrium surface science 
to study the theoretical scalability of 
non-trivial parallelization 
applied to parallel discrete event simulations (PDES), 
such as the dynamic Monte Carlo method.  

\section{Faster Dynamic Metropolis Simulations}

In dynamic Monte Carlo simulations, the dynamic is 
given by the underlying physical system, so it cannot be changed.  
Consequently, 
many of the well-known algorithms, such as loop algorithms, 
cluster algorithms, and multicanonical algorithms cannot be used 
since they are not faithful to the dynamic.  
Furthermore, one Monte Carlo step per spin (MCSS) 
corresponds to an underlying microscopic time \cite{PARK02}, which 
often is much shorter than the time scale 
needed for the simulation.  For example, in simulating ferromagnets 
a Monte Carlo step is approximately an inverse phonon frequency 
\cite{MART77,PARK02}, about $10^{-13}$ seconds.  
The lifetime of a metastable 
state desired for device time scales is years for 
magnetic recording.  In modeling paleomagnetism, the time scales 
of the metastable state are 
millions of years.  To simulate over such disparate time scales 
requires faster-than-real-time algorithms.  

Whenever the rejection rate is high, event-driven rejection-free 
methods are useful.  These include the $n$-fold way \cite{BORT75} 
and its generalization to the MCAMC method \cite{NOVO95}.  
A rejection-free algorithm for 
continuous spin systems has recently been published \cite{MUNO03}.  
An alternative algorithm for first-passage times is the projective 
dynamics method \cite{KOLE98}.  These algorithms can often 
accelerate simulations by many orders of magnitude.  

\begin{table}[!t]
\begin{tabular}{c|c|c|c}
\hline
Spin Orientation& Number of nn spins $\uparrow$ & Small-world spin & 
Flip Probability 
\\
\hline
$  \uparrow$ & 2 & $  \uparrow$ & $p_1$ \\
$  \uparrow$ & 1 & $  \uparrow$ & $p_2$ \\
$  \uparrow$ & 0 & $  \uparrow$ & $p_3$ \\
$  \uparrow$ & 2 & $\downarrow$ & $p_4$ \\
$  \uparrow$ & 1 & $\downarrow$ & $p_5$ \\
$  \uparrow$ & 0 & $\downarrow$ & $p_6$ \\
$\downarrow$ & 2 & $  \uparrow$ & $p_7$ \\
\hline
\end{tabular}
\caption{The spin arrangements for the 
first 7 of 12 spin classes used in the 
MCAMC calculations.  The energies associated with 
these spin configurations enter the spin flip probabilities, $p_i$.}
\end{table}

Here we apply the MCAMC method 
to study metastability of the Ising model on a small-world network.  
The Hamiltonian is 
\begin{equation}
{\cal H} =- J_1 \sum_{i=1}^N \sigma_i \sigma_{i+1} 
          - J_2 \sum_{i=1}^N \sigma_i \sigma_{{\rm sw}(i)} 
          - H   \sum_{i=1}^N \sigma_i
.
\end{equation}
Here $\sigma_i=\pm1$, 
$J_1$ is the ferromagnetic interaction 
along the chain, $J_2$ is a ferromagnetic interaction 
for the small-world connections (see below), 
and $H$ is the applied external field.  
We use periodic boundary conditions for the $N$ Ising spins. 
Each Ising spin has one small-world connection.  
It is obtained by starting with the first spin, and randomly 
connecting it to any of the other $N-1$ spins.  If the next  
spin is not yet connected with a small-world connection, one 
of the remaining unconnected spins is randomly connected to 
it.  These connections are quenched, and do not change in 
a particular simulation.  
Many quenched random small-world bond configurations 
are needed to determine the effect of the randomness.  

The applied Monte Carlo dynamic is: 
1) one of the $N$ spins is chosen at random, 
2) a uniform deviate $r$ on $(0,1]$ is chosen, and 
3) the chosen spin is flipped if the 
Glauber flip probability \cite{LBBook} satisfies
$
r \le \exp(-E_{\rm new}/T)/\left[
\exp(-E_{\rm new}/T)+\exp(-E_{\rm old}/T)\right]
$.  
Here Boltzman's constant has been set to unity, 
$E_{\rm old}$ is the energy of the current spin configuration, and
$E_{\rm new}$ is the energy of the spin configuration with the chosen 
spin flipped.  
We start with all spins $\sigma=+1$, apply a field $H<0$, and 
measure the lifetime $\tau$ until the magnetization is first equal to 
zero.  Using the same quenched random small-world bonds, 
we average over many such escapes to obtain the average lifetime 
$\langle\tau\rangle$ measured in MCSS.  

To measure a metastable lifetime, one 
needs to be below the critical temperature, $T_{\rm c}$.  We estimate 
$T_{\rm c}$ using the Binder fourth-order cumulant 
of the order parameter \cite{LBBook}.  Similar equilibrium studies 
of small-world Ising ferromagnets have recently been performed 
\cite{SWref}.  The crossings of this cumulant 
provide a straightforward way of estimating $T_{\rm c}$ (Fig.~1(a)).  
The average lifetime for $T<T_{\rm c}$ 
grows exponentially in $T^{-1}$ (Fig.~1(b)).  This necessitates 
the use of faster-than-real-time algorithms.  

One way of accelerating the computations is to use a rejection-free 
algorithm.  This includes the $n$-fold way algorithm \cite{BORT75} 
in continuous time, but it also has a counterpart in 
discrete time \cite{NOVO01,NOVO95}.  
When all spins are $+1$, then the probability of flipping 
a single spin in one step is $p_1$ and the average time 
required before a spin flips is $p_1^{-1}$.  Therefore, for small 
$p_1$, computations can be accelerated by asking 
how long it takes to change from the state of all spins up to 
the state with one overturned spin.  This is an example of 
the $s=1$ MCAMC algorithm ($s=1$ transient state, the current state).  
Whenever the spins are all $+1$, the time increment 
$
m = \left\lfloor\
\ln(r_1) / \ln(1-p_1)
\right\rfloor + 1
$
is added, and a randomly chosen spin is flipped.  
Here $r_1$ is a uniformly distributed random number on $(0,1]$, 
$\lfloor\cdot\rfloor$ is the integer part, and 
all spins are equivalent, so we can randomly pick one to flip (in 
the language of the $n$-fold way algorithm, all spins are in the 
same spin class).  

\begin{figure}[tp]
\includegraphics[width=7.0cm]{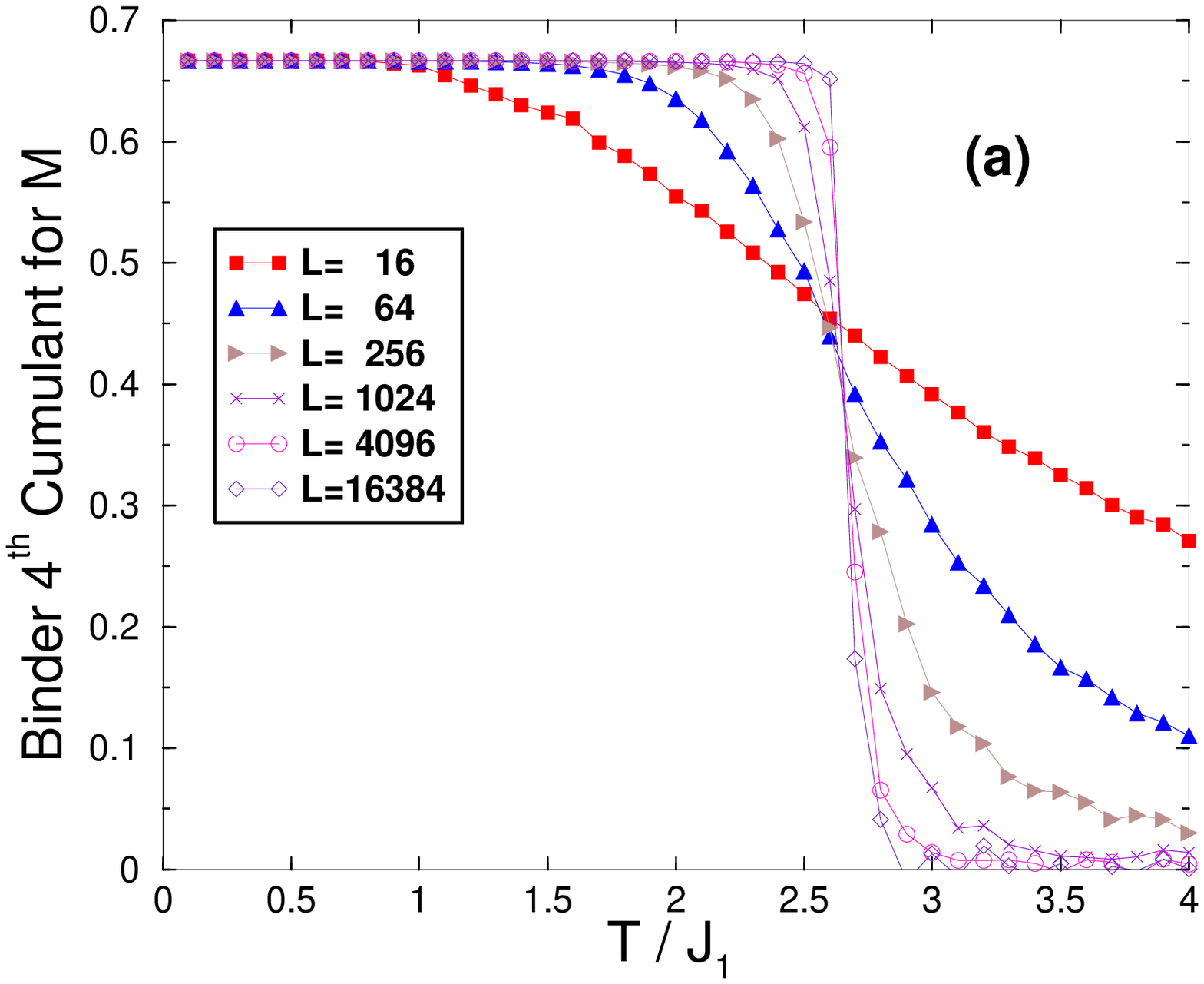}
\includegraphics[width=7.0cm]{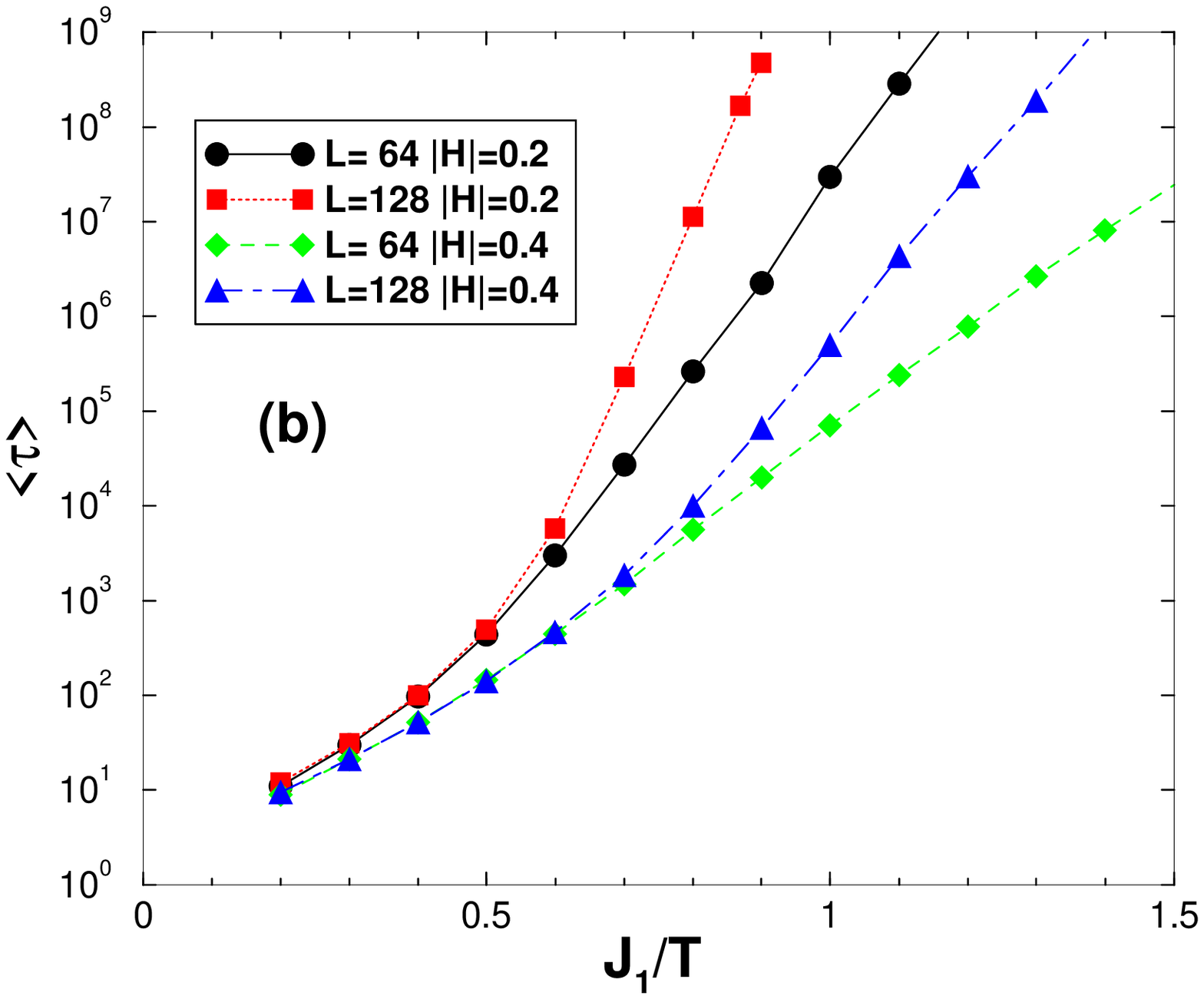}
\caption{\label{FIGINet} 
(a) The Binder fourth-order cumulant for the order parameter for the 
Ising ferromagnet on a small-world network with $H=0$.  The crossings for 
various system sizes gives an estimate of the critical temperature.  
(b) 
The average lifetime in MCSS for $H=-0.2J_1$ and 
$H=-0.4J_1$.  Note the large lifetimes.
}
\end{figure}

At low temperatures and small fields the $s=1$ MCAMC 
algorithm still does not give the best performance.  
However, the performance can be improved 
by adding additional states to the transient subspace.  
For example, for $s=2$ in this 
model, the transient part of the absorbing Markov chain is
\begin{equation}
{\bf T} = 
\pmatrix{
1-x-(p_7/N) & p_7/N \cr
p_1       & 1-p_1 \cr
}
.
\end{equation}
Here $x$ is defined below.  
Then, whenever all spins are $+1$, the time increment 
$m$ that is added to $\tau$, corresponding to exiting to a state 
with two overturned spins, is given 
using a uniform random deviate $r_1$ in $(0,1]$ by 
the solution of 
\begin{equation}
{\vec v}^{\rm T}_{\rm I} {\bf T}^m {\vec e} 
< r_1 \le 
{\vec v}^{\rm T}_{\rm I} {\bf T}^{m-1} {\vec e} 
\end{equation}
where ${\vec e}^{\rm T} =\pmatrix{1 & 1}$ and 
the initial vector is ${\vec v}^{\rm T}_{\rm I}=\pmatrix{0 & 1}$.

Once the time increment $m$ to exit the transient subspace is 
obtained, the next spin configuration must be found, i.e.\ a configuration 
with two overturned spins.  
Let $N_2$ be the number of small-world bonds that connect 
nearest-neighbor (nn) spins.  
Let $x=x_1+x_2$ with 
\begin{eqnarray}
x_1 &= &
\frac{N-N_2}{N^2}
\left[2 p_2 + p_4 + (N-4)p_1\right] 
=
\frac{N-N_2}{N^2} y_1
\\
x_2 & = & 
\frac{N_2}{N^2}
\left[p_5 + p_2 + (N-3)p_1\right] = \frac{N_2}{N^2} y_2
.
\end{eqnarray}
Then the new spin configuration is chosen, 
using uniformly distributed random numbers $r_i$ for $i=2,\cdots,6$.  

If $r_2 x > x_1$, one of the spin pairs with small-world 
bonds longer than nn is randomly chosen using 
$r_3$, one of these two spins is chosen with $r_4$ and is flipped.  
If $r_5 y_1\le 2 p_2$, using $r_6$ one of the two 
nn spins along the chain is chosen and flipped.  
If $2 p_2<r_5 y_1\le 2 p_2 + (N-4) p_1$, the spin connected to the 
flipped spin by the small-world bond is flipped.  
If neither of the two conditions above involving $r_5$ is satisfied, 
then $r_6$ is used to choose one of the other $N-4$ spins (except the 
flipped spin or the 3 spins it is connected to), and the chosen 
spin is flipped.  

If $r_2 x \le x_1$ a similar procedure is used for spins belonging 
to the $N_2$ doubly-connected bonds.  

The MCAMC algorithms do not change the 
dynamics, but rather only implements the dynamics in a 
fashion that enables simulations to longer lifetimes.  
Results for the average lifetime obtained from $10^3$ escapes 
for one realization 
of the quenched small-world bonds are shown in Fig.~1(b).  

\section{Is The Metropolis Dynamic Parallelizable?}

Dynamic Monte Carlo and event-driven rejection-free 
Monte Carlo methods belong to 
a class of problems called discrete-event simulations (DES).  
Non-trivial parallelization of dynamic Monte Carlo and $n$-fold 
way algorithms has been accomplished for Ising spin systems 
\cite{KOLE98,KORN99}.  
Using ideas and methodologies of non-equilibrium surface science, 
it has recently been shown \cite{KORN00} 
that conservative PDES implementations 
should have a virtual time horizon in the 
Kardar-Parisi-Zhang (KPZ) universality class \cite{BS95}.  
Provided that this is the case, then 
{\it all\/} short-ranged 
asynchronous parallel DES simulations can be made to be perfectly scalable. 
This is because, 
as the number of processing elements (PEs) goes to infinity, 
the utilization stays finite \cite{KORN00}, and the measurement portion of 
the algorithm can be bounded \cite{KORN03}.  
A brief review is presented here.  

The stochastic nature of the Metropolis dynamic makes it difficult 
to utilize a parallel computing environment to the fullest 
extent because {\it a priori} there is no global clock to 
synchronize physical processes in a system with asynchronous 
dynamics. However, the system is not inherently serial. 

The methodology for PDES simulations works in all 
dimensions, but for simplicity we consider 
parallelization of dynamic Monte Carlo for a one-dimensional Ising model.  
In non-trivial parallelization, the spin system is spatially 
distributed among $L$ processing elements, i.e., physical 
processes and interactions between physical subsystems are mapped 
to logical processes and logical dependences between PEs 
(Fig.~\ref{chart}). 
In our model of PDES performance for the spin system with 
nn interactions, we consider an ideal system of $L$ identical 
PEs, arranged on a ring, where communications between PEs take place 
instantaneously. 
Each PE manages the state of the assigned 
subsystem of $N$ spins, and has its own time 
(called the local virtual time, LVT).  
The LVT progresses on each PE during the simulation. 
The asynchronous nature of physical dynamics implies an asynchronous 
system of logical processes. 
Logical processes execute concurrently 
and exchange time-stamped messages to perform state updates of 
the entire physical system being simulated. A sufficient condition 
for preserving causality in simulations requires that each logical 
process works out the received messages from other logical processes 
in non-decreasing time-stamp order \cite{JEFF85,FUJI90}. 
PDES are classified 
in two categories: optimistic \cite{JEFF85} and conservative 
\cite{LUBA8X,CM79,Mis86}. In conservative PDES, an algorithm does 
not allow a logical process to advance its LVT (i.e., to proceed with 
computations) until it is certain that no causality violation can occur. 
In optimistic PDES, an algorithm allows a logical process to advance 
its LVT regardless of the possibility of a causality error. The optimistic 
scenario detects causality errors and provides a recovery procedure 
to detect and fix such errors. 
Several aspects of a PDES algorithm should be considered in 
efficiency studies, including: 
the synchronization procedures; 
the average utilization $\langle u \rangle$ of the parallel environment 
as measured by the mean fraction of working PEs between update attempts; 
the memory requirements per PE; 
and 
the scalability as measured by evaluating 
the performance when $L$ is increased.  

\begin{figure}[bp]
\includegraphics[width=8.63cm]{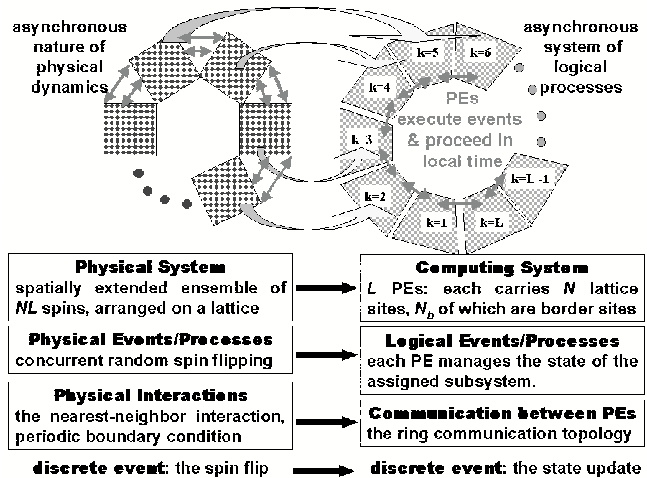}
\caption{\label{chart} The mapping of short-ranged 
physical processes to logical processes. 
The nn physical interactions (two-sided arrows 
in the left part) on a lattice with  periodic boundary conditions 
are mapped to the ring communication topology of logical processes 
(two-sided arrows in the right part). Each PE carries $N$ lattice sites, 
but communications take place only for border sites.}
\end{figure}

In our study the main concept is the virtual time horizon (VTH), 
defined as the set of the LVTs for all logical processes. 
We model the 
growth of the VTH as a deposition process of Poisson random time 
increments on a one-dimensional lattice of $L$ processors.   
The growth rule of the VTH is defined by the PDES algorithm.  
The width of the VTH provides a measure of 
the desynchronization in the system of PEs and is related to the 
memory requirements for parallel simulations \cite{KORN03,KNKG02,KNK03}. 
Here the principle is: the larger the width, the larger the memory required 
per PE. The asymptotic scalability of an 
algorithm can be assessed by 
applying coarse-grained methods to the VTH \cite{KORN00}. Computational 
speed-up (as measured by comparing the performance of the parallel 
with sequential simulations) can be derived from the microscopic 
structure of the VTH \cite{KRK03}. 

In modeling a conservative PDES, at each update attempt $t$, on each PE 
the simulation algorithm randomly selects one of the $N$ spin sites. 
If the 
selected site is an interior site, the update happens and the simulated 
LVT is incremented for the next update attempt: 
$\tau (t+1)= \tau(t)+ \eta$, where $\eta$ is a random time increment 
that is sampled from the Poisson distribution with unit mean. 
If the selected site is a border site, 
the PE must {\it wait until\/} the LVT of its neighbor(s) is not less than 
its own LVT, at which time the waiting PE makes the update and proceeds.  
For $N=1$ the LVT of both neighboring PEs are considered, while for 
$N>1$ only the corresponding neighboring PE's LVT is considered.   

In the most unfavorable case of conservative parallelization $N=1$. 
For  such a closed 
spin chain the mean utilization $\langle u(L;N=1) \rangle$ of the 
parallel processing environment is simply the mean density of local 
minima in the conservative VTH during the steady state.  
Analyzing the microscopic structure of the 
VTH at saturation, it is possible to derive approximate analytical 
formulas for $\langle u(L;N) \rangle$ and the higher moments of 
$u(L;N)$ (Fig.~3(a)).  
For example, 
\begin{equation}
\label{umom}
  \begin{array}{rclcl} 
  \langle u(L;1)\rangle &=& (L+1)/4L ~,  &~~~~~& L\ge 3 \\ 
  \langle u(L;2)\rangle &=& (3L+1)/8L ~, &~~~~~& L\ge 3 \\
 \end{array} 
\, .
\end{equation}
Note that as $L\rightarrow\infty$, the utilization is 
about 1/4 for $N=1$ and about 3/8 for $N=2$.  For large $N$, the 
asymptotic utilization can be near the theoretical limit of unity.  

The conservative PDES utilization depends on $N$, 
as well as on the number $N_b$ of effective 
border lattice sites per PE (here $N_b=2$), 
and on the communication topology. 
Our earlier large-scale simulations \cite{KNK03} 
show that the worst-case ($N=1$) conservative scenario for a spin chain can 
be greatly improved when $N$ is increased while retaining the ring 
communication topology with $N_b=2$ (Fig.~\ref{FIGpdes}(b)). 
Thus, to take the best advantage of 
conservative parallelization one should use 
many PEs with many spins per PE 
[see Fig.~\ref{FIGpdes}(b)]. 
In this case, preliminary analysis of the width of the VTH 
shows it scales as: 
\begin{equation}
\label{width}
\langle w (t) \rangle \sim  \left\{ { \begin{array}{cl}
t^{\beta _1} & ,\>\> t \ll t_{1 \times} \\ 
t^{\beta _2} & ,\>\> t_{1 \times} \ll t \ll t_{2 \times} \\ 
L^\alpha \sqrt{N} & ,\>\> t \gg t_{2 \times}
 \end{array} } \right. ,
\end{equation}
where $\alpha=1/2$ and the cross-over times are: 
$t_{1 \times} \sim  N$ independent of $L$; 
$t_{2 \times} \sim N L^{z}$.  
Here $z= \alpha / \beta_2$ is the 
dynamic exponent, and the growth 
exponents are $\beta_1\approx1/2$ (corresponding to random deposition) 
and $\beta_2\approx1/3$ (corresponding to the KPZ 
universality class) \cite{BS95}. 
The scaling exponent $\alpha=1/2$ of the VTH width at saturation 
(for $t \gg t_{2 \times}$) implies that the memory requirement for the state 
savings grows as a power law, 
i.e., as $\sqrt{LN}$. 
Recent applications of the conservative algorithm to modeling magnetization 
switching \cite{KORN99} and a dynamic phase transition in highly anisotropic 
thin-film ferromagnets \cite{KWRN01,KRN02} indicate that conservative 
parallelization can be very efficient in simulating spin dynamics with 
short-range interactions.

\begin{figure}[tp]
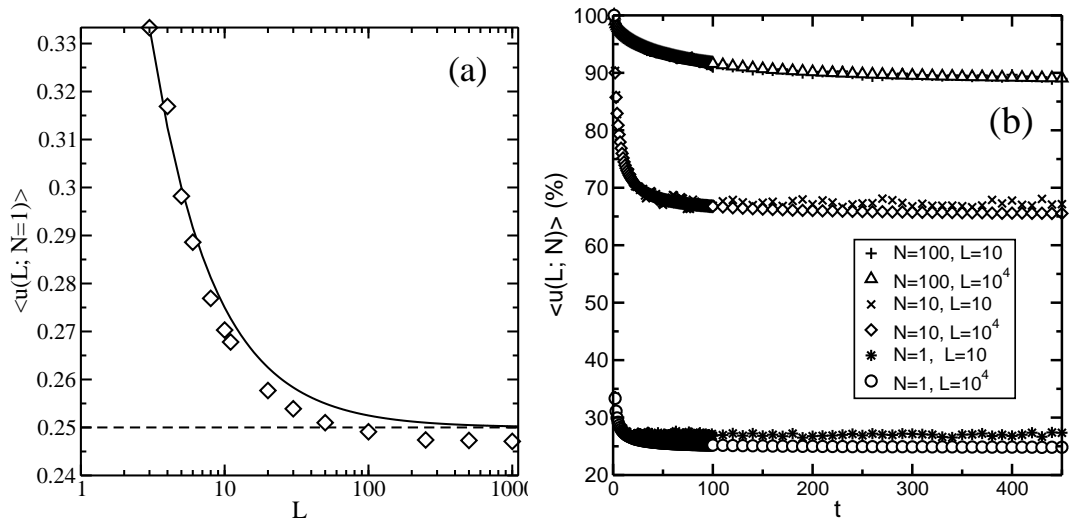

\includegraphics[width=7.0cm]{metropolis-fig04a.eps}
\includegraphics[width=7.0cm]{metropolis-fig05b.eps}
\caption{\label{FIGpdes} 
(a) The steady-state mean utilization vs 
the system size in conservative PDES for a spin chain 
with $N=1$. Analytical 
result (solid curve); 
infinite-$L$ limit (dashed line); and simulation data (symbols).
(b) The time evolution of the mean utilization 
in conservative PDES for spin chains 
when each PE carries $N$ spin sites, two of which are the effective border 
sites. Observe that the utilization grows with $N$.}
\end{figure}

\section{Discussion and Conclusions}

This brief paper has described how to make dynamic Monte Carlo 
simulations faster and larger.  The algorithms described do 
not change the dynamics in any fashion, but rather implement 
the dynamics on the computers using advanced techniques.  

To accelerate the simulations, faster-than-real-time algorithms 
may be implemented.  These include the $n$-fold way algorithm 
\cite{BORT75} and its extension, the Monte Carlo with Absorbing Markov 
Chain (MCAMC) algorithm \cite{NOVO01,NOVO95}, as well as the projective 
dynamics method \cite{KOLE98}.  We outlined 
$s=1$ and $s=2$ MCAMC methods, as applied 
to magnetic field-reversal in 
a ferromagnetic Ising model on a small-world network.  

To make the simulations larger, non-trivial parallelization is 
required.  We briefly described how ideas from non-equilibrium 
surface science 
can be used to understand such simulations.  In particular, all short-ranged 
conservative PDES 
should have a virtual time horizon governed by the KPZ universality class.  
In this case, {\it all\/} short-ranged 
PDES (such as 
dynamic Monte Carlo) can be made scalable using a 
conservative PDES approach.  
Conservative PDES references include 
\cite{NOVO01,LUBA8X,KORN99,KORN00,KORN03}.  
The alternative implementation, optimistic PDES simulations 
\cite{JEFF85,FUJI90} 
for dynamic Monte Carlo simulations, 
have shown some of the difficulties with scalability \cite{SOS01}.  


\begin{theacknowledgments}
Supported in part by NSF grants DMR-0113049 and DMR-0120310.  
\end{theacknowledgments}


\bibliographystyle{aipprocl} 



\end{document}

\endinput